\documentclass[12pt]{iopart}
\usepackage[utf8]{inputenc}
\usepackage{graphicx}
\usepackage{psfrag}
\usepackage{iopams}

\def\correction#1{#1}       % Final version

\def\be{\begin{equation}}
\def\ee{\end{equation}}

\usepackage[dvipsnames]{xcolor}

\begin{document}
\title{CTMRG study of the critical behavior of an interacting-dimer model}
\author{C Chatelain}
\address{Universit\'e de Lorraine, CNRS, LPCT, F-54000 Nancy, France}
\ead{christophe.chatelain@univ-lorraine.fr}
\date{\today}

\begin{abstract}
The critical behavior of a dimer model with an interaction favoring
parallel dimers in each plaquette of the square lattice
is studied numerically by means of the Corner Transfer Matrix
Renormalization Group algorithm. The critical exponents are known to
depend on the chemical potential of vacancies, or monomers. At large
average density of the latter, the phase transition becomes of first-order.
We compute the scaling dimensions of both order parameter and
temperature in the second order regime and compare them with the
conjecture that the critical behavior is the same as the Ashkin-Teller
model on its self-dual critical line.
\end{abstract}
\maketitle

\section{Introduction}
Besides their experimental realizations, in particular as diatomic molecules
adsorbed on a surface~\cite{Fowler,Roberts}, dimer models, are, with
the Ising model and its generalizations, among the most studied toy
models of Statistical Physics. They are also encountered in the ground state
of other models, for instance the fully-frustrated Ising model on a
triangular or square lattice~\cite{Nienhuis} or in Resonant
Valence Bond states of quantum spin-$1/2$ antiferromagnets~\cite{Anderson,Lacroix}.
In the close-packed limit, the number of coverings of a 2D lattice by
dimers has been computed exactly by Temperley, Fisher, and Kasteleyn~%
\cite{Temperley,Fisher,Kasteleyn}. Due to the intrinsic geometric frustration
of the model, the entropy per site is non-zero and dimer-dimer
correlation functions decay algebraically as $1/r^2$ with the distance $r$.
When two lattice sites are left empty or, equivalently, when
they are occupied by a monomer, the number of possible dimer configurations
decay algebraically with the distance $r$ separating the monomers as
$1/\sqrt r$~\cite{Stephenson,Hartwig}. The case of a non-zero density
of monomers has also been considered. The density can be either fixed or
controled by a chemical potential $\mu_1$. It has been shown that this
monomer-dimer model does not undergo any temperature or density-driven
phase transition~\cite{Heilmann1,Heilmann2}.
\\

\begin{figure}
    \centering
    \includegraphics[width=0.54\textwidth]{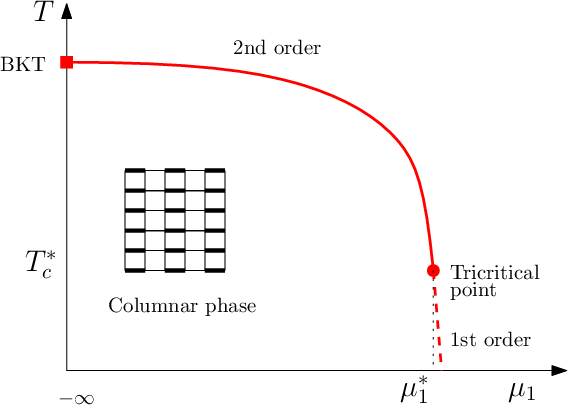}
    \caption{
    Schematic phase diagram of the monomer-dimer model
    with an interaction between parallel dimers in a plaquette
    (figure inspired by Fig.~1 of Ref.~\cite{Alet2}).
    The close-packed limit, corresponding to the limit
    $\mu_1\rightarrow -\infty$, is represented on the $y$-axis.
    $T_c^*$ and $\mu_1^*$ are respectively the temperature and
    the monomer chemical potential at the tricritical point.
    }\label{fig0}
\end{figure}

Nevertheless, a phase transition can be observed when an interaction is
introduced between neighboring dimers. The case of an interaction favoring
parallel dimers in a plaquette of the square lattice has been studied in
detail~\cite{Alet1,Alet2,Li,Roychow,Morita}. In the close-packed limit, Monte Carlo
simulations showed that the system undergoes a Berezinskii-Kosterlitz-Thouless
(BKT) phase transition at $T_{\rm BKT}=0.65(1)$ between a low-temperature
columnar ordered phase and the critical dimer phase at high temperature
(Fig.~\ref{fig0}). For a finite monomer chemical potential
$\mu_1<\mu_1^*$, and therefore a non-zero monomer density, the system
undergoes a continuous phase transition with critical exponents varying
with $\mu_1$. Transfer matrix estimates of the four smallest scaling
dimensions by the gap-exponent relation support a Coulomb gas
picture which implies that the interacting-dimer model shares the same
critical behavior as the isotropic Ashkin-Teller model along its self-dual
line. Thanks to a mapping of the latter onto the 8-vertex model, its
critical exponents have been shown to be~\cite{Kadanoff,Nienhuis2,Baxter}
    \be x_{\sigma}={1\over 8},\quad
    x_{\sigma\tau}={1\over 8-4y},\quad
    y_t={3-2y\over 2-y}         \label{ScalDimAT}\ee
where the parameter $y$ is in the range $[0;3/2]$ for the
Ashkin-Teller model.
The model is equivalent to the 4-state Potts model when $y=0$ and to two
decoupled Ising models when $y=1$. The interacting-dimer model in the
close-packed limit $\mu_1\rightarrow -\infty$ has been conjectured to
correspond to $y=3/2$. For $0\le y\le 3/2$, the scaling dimension of the
dimer operator coupled by the interaction is given by $x_{\sigma\tau}$.
At the monomer chemical potential $\mu_1^*$, corresponding to
$y=0$, the phase transition becomes first-order, as predicted by
mean-field theory~\cite{Alberici}.
Like the interacting-dimer model, a mixture of hard squares and dimers
have been shown to undergo a phase transition that belongs to the
Ashkin-Teller universality class~\cite{Ramola}.
The cases of anisotropic dimer interactions and repulsive interactions
have been considered by transfer matrix calculations~\cite{Otsuka,Otsuka2}.
On a cubic lattice, Monte Carlo simulations showed that the transition is
continuous only in presence of competing plaquette and cubic interactions
and a tricritical point~\footnote{
    The word {\sl tricritical} is used here to denote the end of
    the transition line of second order. Since more than 3 phases
    are in coexistence at the first-order phase transition, the
    word {\sl multicritical} would be more rigourous.}
is also observed~\cite{Alet3}.

In the last decade, various numerical Tensor Network (TN) algorithms
for the interacting-dimer model have been considered as an alternative to
Monte Carlo simulations and transfer matrices. TN calculations are
expected to converge faster than Monte Carlo simulations, even with
worm algorithm, and to give access to much larger lattice sizes than
transfer matrices. However, TN techniques are also known to be efficient
only for gapped systems, i.e. away from any critical point. The
unavoidable truncation of the tensors induces systematic deviations
close to critical points that are difficult to estimate.
It is therefore useful to test different TN algorithms and
different data analysis methods to estimate critical exponents.
Li {\sl et al.} studied the interacting-dimer model by contracting
the TN using an infinite time evolving block decimation (iTEBD) algorithm~\cite{Li}
and confirmed the phase diagram of Ref.~\cite{Alet2}. The algebraic decay
of correlation functions has been reproduced by Tensor Renormalization
Group~\cite{Roychow}. Recently, the Ashkin-Teller universality class has
been tested by both TN methods and Monte Carlo simulations~\cite{Morita}.
In this paper, we present a study of the interacting-dimer model by
means of a particular TN algorithm, the Corner Transfer Matrix
Renormalization Group (CTMRG)~\cite{Nishino1, Nishino2, Nishino3}.
The latter is an extension of the celebrated Density Matrix Renormalization
Group (DMRG) algorithm~\cite{White1, White2} to classical statistical-mechanics
models. It has been applied to a variety of lattice spin systems:
Ising model in the hyperbolic plane~\cite{Gendiar1,Gendiar2},
clock model~\cite{Gendiar3,Gendiar4,Gendiar5}, chiral Ashkin-Teller
model~\cite{Mila}, vertex model~\cite{Gendiar6} but also hard
squares~\cite{Mila2} and hard rods~\cite{Chatelain}. The CTMRG algorithm
is also used to contract infinite Projected Entanglement-Pair States
(iPEPS) that provide an efficient representation of 2D quantum
states~\cite{Orus,Cirac}.
\\

The plan of this paper is the following: the model and the algorithm are detailed
in the first section. Our estimates of the transition temperatures are presented
in section~\ref{sec2} and compared with those obtained in Ref.~\cite{Alet2} by transfer
matrix calculations. In section~\ref{sec3}, the order-parameter and temperature scaling
dimensions are computed along the transition line in the regime of second-order
phase transition. The location of the tricritical point is inferred and the
conjecture of an Ashkin-Teller universality class is tested.
Conclusions follow.

\section{Model and CTMRG algorithm}\label{sec1}
\subsection{The interacting-dimer model}\label{sec1.1}
We consider a square lattice $\Lambda=(E,V)$ where $V$ denotes the set of vertices
of the lattice and $E\subseteq V\times V$ the set of edges between nearest
neighboring vertices. Dimers occupy edges of the lattice $\Lambda$ and are not allowed
to overlap, i.e. if the edge $(i,j)$ is occupied by a dimer, no dimer can be
found on the edges $(i,k)$ and $(k,j)$ of $E$. In the following, we introduce
a variable $n_{ij}=n_{ji}$ equal to 1 if a dimer is present on the edge $(i,j)$
and 0 otherwise. The vertices that are not covered by any dimer are considered
to be occupied by a monomer. The variable $n_i$ is set to 1 if a monomer is
present on the vertex $i$ and 0 otherwise. Note that the $n_i$'s are not
independent variables: $n_i=\prod_k (1-n_{ik})$ where the product extends over
the neighbors $k$ of the vertex $i$.
\\

The average density of monomers is fixed by
a chemical potential $\mu_1$. The presence of a monomer is therefore affected
by a statistical weight $z=e^{\beta\mu_1}$ where $\beta=1/k_BT$ ($k_B=1$)
is the inverse temperature.
In addition, an interaction is introduced to favor the presence of dimers
on two parallel edges of the same plaquette. The Boltzmann weight of a
dimer configuration $\{n_{ij}\}$ is finally
    \be W(n_{ij})=e^{-\beta[uH+\mu_1N_1]}\label{PoidsSt}\ee
where
    \be N_1=\sum_i n_i\ee
is the number of monomers and
    \be H=\sum_{(i_1,i_2,i_3,i_4)\in\opensquare}
    \big[n_{i_1i_2}n_{i_3i_4}+n_{i_1i_3}n_{i_2i_4}\big]\ee
is the interaction energy. The symbol
$\opensquare\subseteq V^{\times 4}$ denotes the set
of plaquettes of the lattice that are formed by the edges $\{i_1,i_2,i_3,i_4\}$
with $(i_1,i_2)$ and $(i_3,i_4)$ being horizontal edges while $(i_1,i_3)$
and $(i_2,i_4)$ being vertical ones. In the following, the coupling constant
$u$ will be chosen equal to 1, without loss of generality.
\\

Numerical calculations have shown that the low temperature phase
is a columnar phase where dimers are all in the same direction,
either horizontal or vertical, in every second columns or
lines~\cite{Alet1,Alet2}.
The ground state is therefore four-fold degenerated and breaks two
${\mathbb{Z}_2}$ symmetries: the rotation of the lattice by $90^\circ$
and the discrete translation by one lattice step. These
two symmetries are simultaneously broken at the phase transition. Since
TN algorithms are more efficient away from critical points, we introduced
a small field breaking the rotational symmetry in two different ways. First,
we considered a small dimer chemical potential $\Delta\mu$ with an opposite
sign for horizontal and vertical dimers:
    \be W(n_{ij})=e^{-\beta[uH+\Delta\mu N+\mu_1N_1]}
    \label{PoidsSt2}\ee
where
    \be N=\sum_{(i,j)\in E_H} n_{ij}-\sum_{(i,j)\in E_V} n_{ij}\ee
is the difference between the number of horizontal dimers and the number
of vertical dimers. We also considered a small shift of the interaction
coupling $u$ with different signs for plaquettes with parallel
horizontal dimers and with vertical ones:
    \be W(n_{ij})=e^{-\beta[uH+\Delta u P+\mu_1N_1]}
    \label{PoidsSt3}\ee
where
    \be P=\sum_{(i_1,i_2,i_3,i_4)\in\opensquare}
    \big[n_{i_1i_2}n_{i_3i_4}-n_{i_1i_3}n_{i_2i_4}\big]\ee
is the difference between the number of plaquettes with two horizontal
dimers and the number of plaquettes with two vertical ones. The quantities
$P$ and $N$ are not equivalent, but \correction{both} break the same symmetry.
\correction{Even though it is not obvious that they have the same scaling
dimension, it will be shown in the following that it is indeed the case.}
They will \correction{therefore} provide two independent estimates of the
critical exponents.

\begin{figure}
    \centering
    \includegraphics[width=0.45\textwidth]{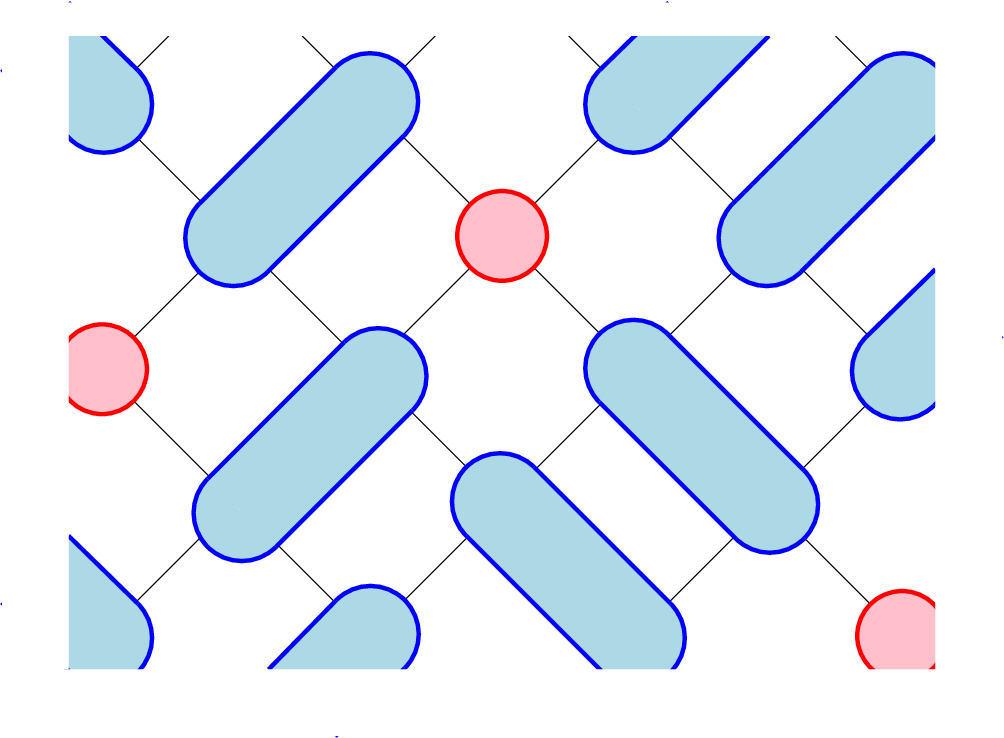}
    \caption{Example of a monomer-dimer configuration on the square
    lattice. Monomers are represented as red circles centered on lattice
    sites while dimers, in blue, overlap edges.}
    \label{fig1}
\end{figure}

\begin{figure}
    \centering
    \includegraphics[width=0.45\textwidth]{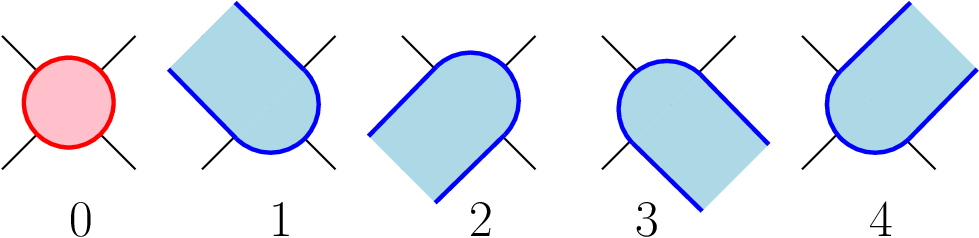}
    \caption{Indices associated to the five possible states of the system
    on a vertex of the lattice.}
    \label{fig2}
\end{figure}

\begin{figure}
    \centering
    \includegraphics[width=0.45\textwidth]{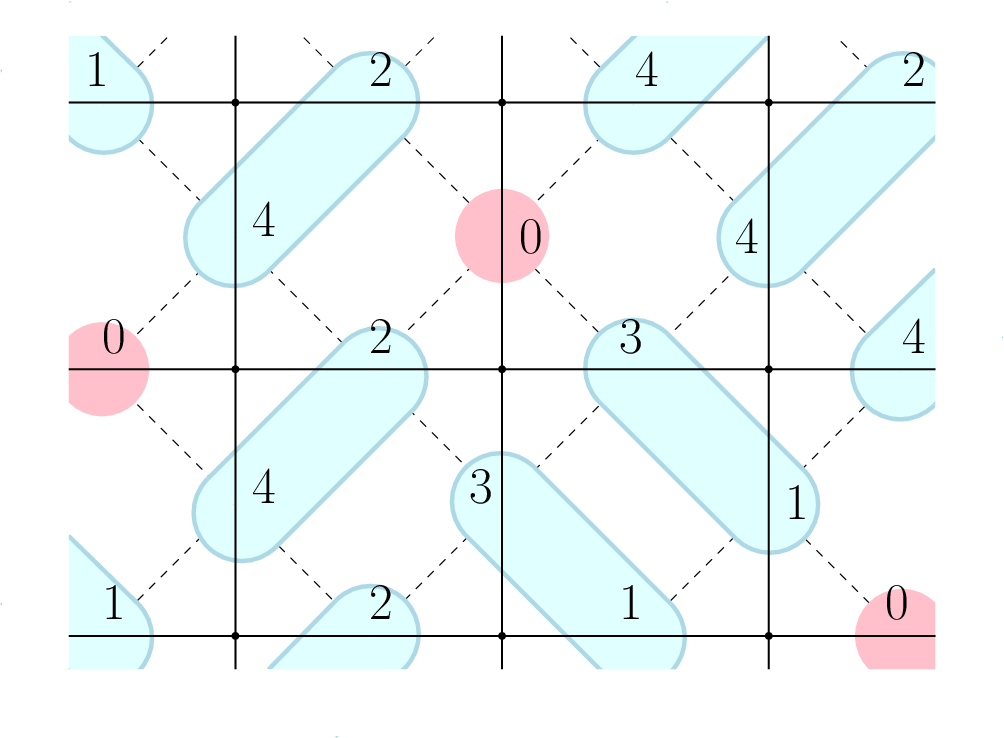}
    \caption{Vertex indices $s_i$ corresponding to the monomer-dimer configuration
    of Fig.~\ref{fig1}. The original lattice $\Lambda$ is represented as dashed
    lines and the new lattice $\tilde\Lambda$ as continuous lines.}
    \label{fig3}
\end{figure}

\begin{figure}
    \centering
    \includegraphics[width=0.45\textwidth]{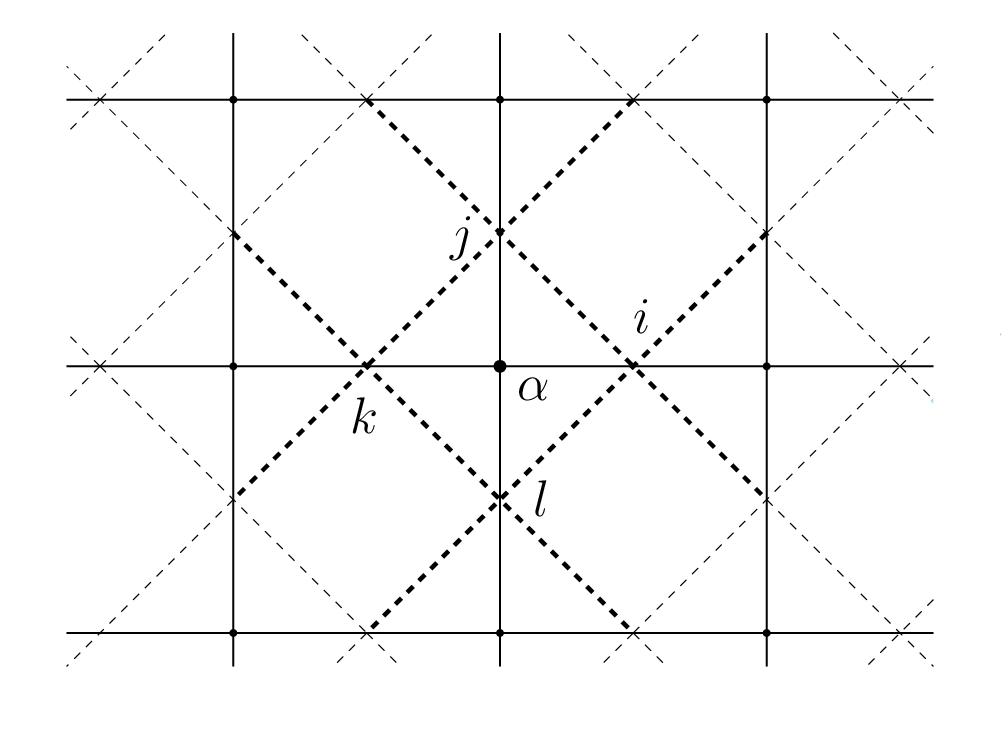}
    \caption{
    As in Fig.~\ref{fig3}, the original lattice $\Lambda$ is represented
    as dashed lines and the new lattice $\tilde\Lambda$ as continuous lines.
    A four-rank tensor $w_{s_is_js_ks_l}$ is placed at each vertex of
    $\tilde\Lambda$. For the vertex $\alpha$ (black disk at the center
    of the figure), the four indices $s_i,s_j,s_k$ and $s_l$ are located
    on the four nearest sites $i,j,k$ and $l$ of $\Lambda$. The index
    $s_i$ (resp. $s_j$, $s_k$ and $s_l$) encodes the monomer-dimer
    configuration on the four edges of $\Lambda$ linked to $i$
    (resp. $j,k$ and $l$). The knowledge of $s_i,s_j,s_k$ and $s_l$
    allows to reconstruct the dimer occupancy of the 12 edges of
    $\Lambda$ represented by bold dashed lines on the figure.}
    \label{fig3b}
\end{figure}

\subsection{CTMRG algorithm for the dimer model}
To be able to use the Corner Transfer Matrix Renormalization Group (CTMRG)
algorithm, the interacting-dimer model was first reformulated.
Any monomer-dimer configuration $\{n_{ij}\}$ can be uniquely characterized
by a set of indices $s_{i}\in\{0,1,2,3,4\}$ defined on the vertices of the
lattice $\Lambda$. An example of such a possible one-to-one map is shown
on Fig.~\ref{fig2}. The monomer-dimer configuration of Fig.~\ref{fig1} for
instance is encoded in the indices $s_i$ of Fig.~\ref{fig3}.
The indices $s_i$ always lead to a monomer-dimer configuration
satisfying the constraint that two dimers cannot overlap.
However, the indices $s_i$ are not independent variables and should satisfy
some compatibility constraints. In particular, if a dimer is present on
the edge $(i,j)$ then $s_i$ and $s_j$ should be equal to either $1$ and
$3$ or $2$ and $4$, depending on the orientation of the edge.
We consider then the square lattice $\tilde\Lambda=(\tilde V,\tilde E)$,
at $45^\circ$ of the original one, for which the vertices $V$
of $\Lambda$ lay at the center of the edges $\tilde E$ of $\tilde\Lambda$.
The indices $\{s_i\}$ are therefore carried by the edges of $\tilde\Lambda$.
One can show that the statistical weights Eq. \ref{PoidsSt}, \ref{PoidsSt2},
and \ref{PoidsSt3} can be decomposed into a product of tensors:
    \be W(s_i)=\prod_{\alpha\in\tilde V} \prod_{i,j,k,l
    \in\tilde E_{\alpha}} w_{s_is_js_ks_l}.\ee
A tensor $w$ of rank 4 is found at each vertex $\alpha\in\tilde V$
of the lattice $\tilde\Lambda$. The set $\tilde E_{\alpha}\subset \tilde E$
is the subset of the edges connected to the vertex $\alpha$.
From the indices $s_i$, $s_j$, $s_k$ and $s_l$, one can reconstruct the
occupancy by a monomer on the 4 vertices $i,j,k,$ and $l$ of the original
lattice $\Lambda$ that are located around $\alpha$, and the occupancy by
a dimer on 12 edges (see Fig.~\ref{fig3b}). Therefore, the indices
$s_i$, $s_j$, $s_k$ and $s_l$ give \correction{complete} information
on the presence of aligned dimers, either horizontal or vertical,
\correction{on these 12 edges} around $\alpha$.
\\

Since each index can take 5 values, the tensor $w$ has $5^4$ entries.
However, most of them should always be equal
to zero to impose the compatibility constraints between the indices
$s_i$, $s_j$, $s_k$, $s_l$.
Our tensor decomposition is different from the one introduced by Baxter
for the monomer-dimer problem~\cite{Baxter2}. Baxter's tensors have only
$2^4$ entries but do not contain any information on the presence of
aligned dimers in a given plaquette and therefore cannot be used to describe
interacting dimers. Our tensors are more lightweight than the
one employed in Ref.~\cite{Li} which have $7^4$ entries, compared to $5^4$
for ours. Since the CTMRG algorithm involves Singular-Value
Decompositions, we expect a smaller error with our formulation after
the truncation to the same number of Singular Values.
The tensors of Ref.~\cite{Morita} have $5^4$ entries too
but they lay on the original lattice $\Lambda$. Our decomposition involves
therefore fewer tensors for the same number of sites.

\begin{figure}
    \centering
    \includegraphics[width=0.52\textwidth]{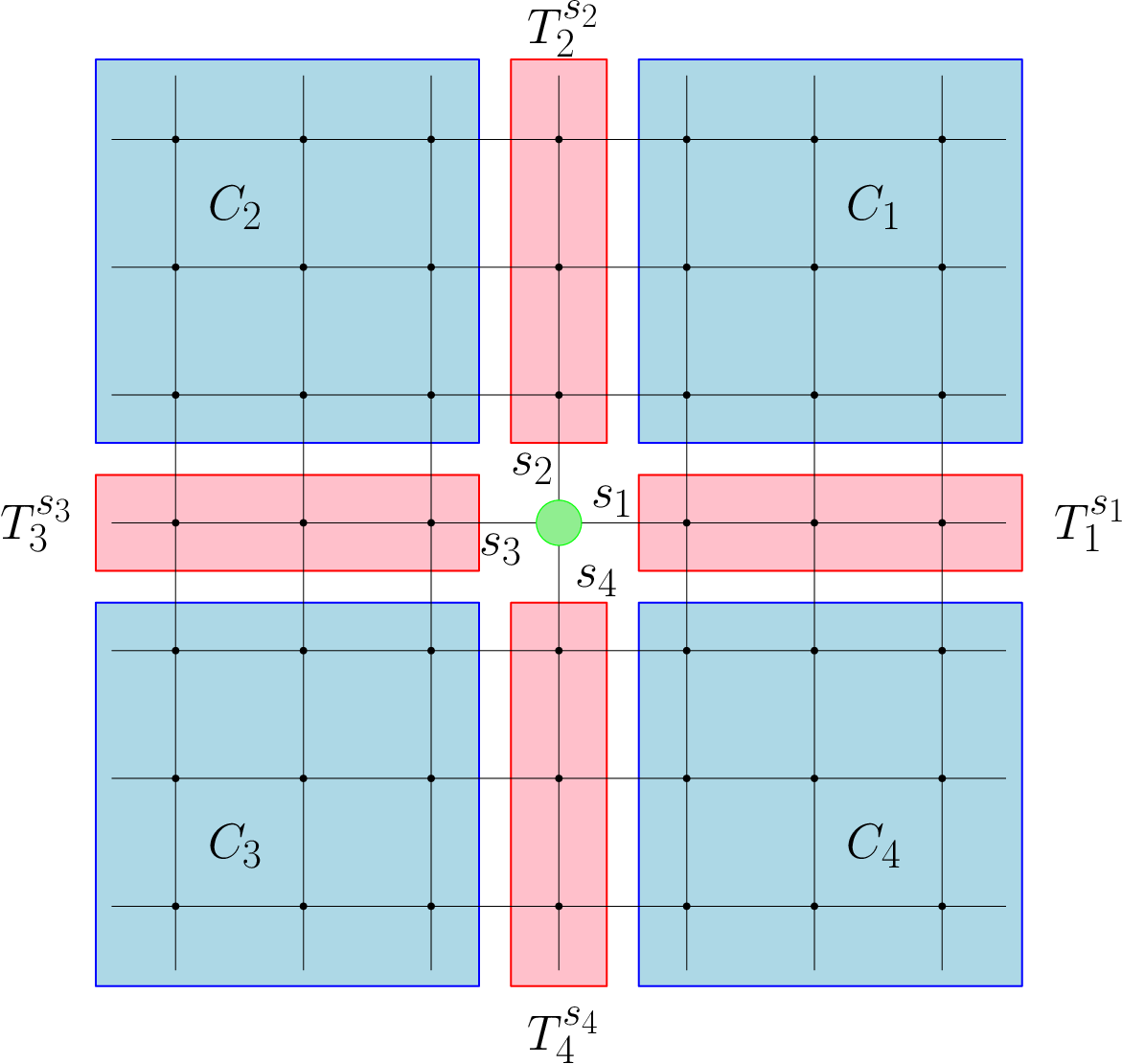}
    \caption{Decomposition of the partition function
    of a $7\times 7$ system into a
    central vertex (green), four Transfer matrices (red) and four
    Corner Transfer matrices (blue). The indices of the
    various matrices are the indices $s_i$ introduced above.
    They are on the edges of $\tilde\Lambda$ which are represented
    as continuous lines on the figure.}
    \label{fig4}
\end{figure}

\begin{figure}
    \centering
    \includegraphics[width=0.39\textwidth]{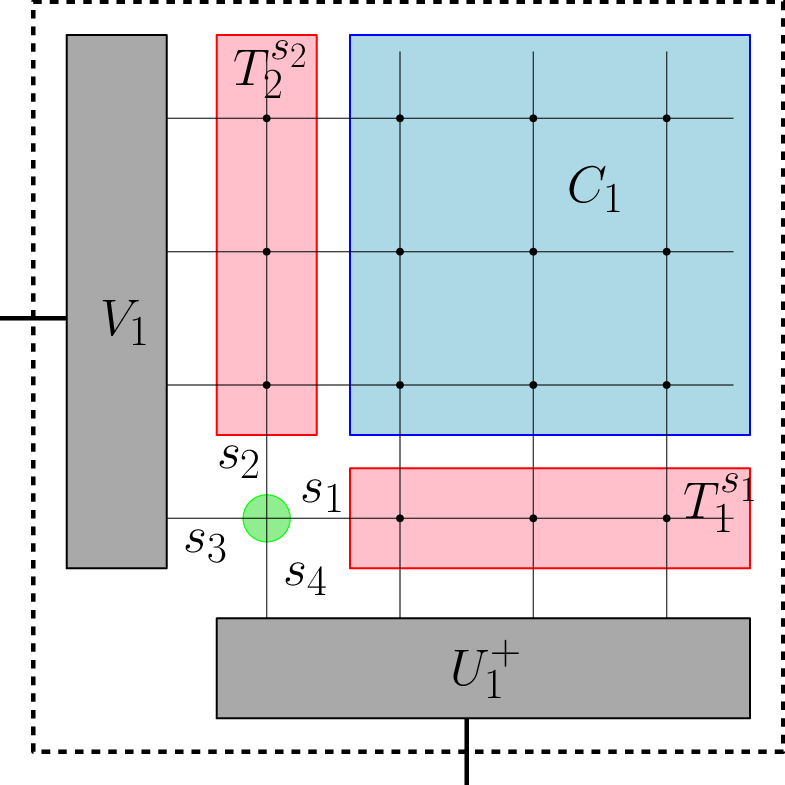}
    \caption{Extension and renormalization of the Corner Transfer
    matrix $C_1$. The matrices $U_1$ and $V_1$ are rectangular matrices
    obtained from the truncation of the unitary matrices that decompose
    the extended Corner Transfer matrix into singular values.}
    \label{fig5}
\end{figure}

\begin{figure}
    \centering
    \includegraphics[width=0.25\textwidth]{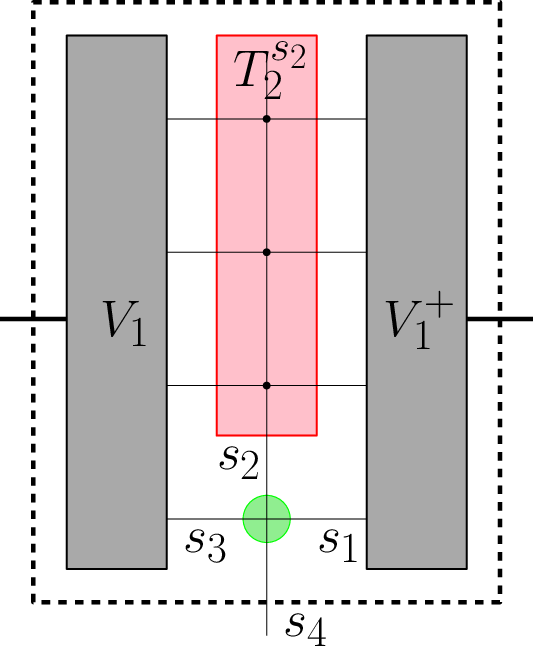}
    \caption{Extension and renormalization of the Transfer matrix $T_2$.}
    \label{fig6}
\end{figure}

The system is studied numerically by means of the CTMRG
algorithm~\cite{Nishino1,Nishino2,Nishino3}.
The partition function of the system is decomposed into a product
    \be{\cal Z}=\sum_{s_1,s_2,s_3,s_4} w_{s_1s_2s_3s_4}\Tr\big[
    T_1^{s_1}C_1T_2^{s_2}C_2T_3^{s_3}C_3T_4^{s_4}C_4\big]
    \label{Part}\ee
where the $T_i^{s}$ are the four Transfer matrices with a boundary degree
of freedom $s$ and the $C_i$ are the four Corner Transfer matrices
(Fig.~\ref{fig4}). Averages of local observables are estimated as
    \begin{eqnarray}
    \langle O\rangle={1\over{\cal Z}}&&\sum_{s_1,s_2,s_3,s_4}
    w_{s_1s_2s_3s_4}O_{s_1s_2s_3s_4}            \nonumber\\
    &&\times\Tr\big[T_1^{s_1}C_1T_2^{s_2}C_2T_3^{s_3}C_3T_4^{s_4}C_4\big].
    \end{eqnarray}
The algorithm starts with a $3\times 3$ system with Free Boundary Conditions.
Each one of the four matrices $T_i^s$ and $C_i$ is attached to one site of
the lattice $\tilde\Lambda$. After imposing the Boundary Conditions, their
matrix elements are given by contractions of the tensor $w$, for example
    \be [T_2^{s_4}]_{s_1,s_3}=\sum_{s_2} w_{s_1s_2s_3s_4},\hskip 1cm
    [C_1]_{s_4,s_3}=\sum_{s_1,s_2} w_{s_1s_2s_3s_4}.    \label{InitCTM}\ee
The order of the indices of $w$ corresponds to the convention
shown on Fig.~\ref{fig3b}.
The first step of the CTMRG algorithm consists in the extension of the
four Corner Transfer matrices and Transfer matrices by the addition of a
new vertex as depicted on Fig.~\ref{fig5} and \ref{fig6}.
For the Transfer matrix, the extension reads
    \be [{T'}_2^{s_4}]_{(s_1s_5),(s_3s_6)}
    =\sum_{s_2} w_{s_1s_2s_3s_4}[T_2^{s_2}]_{s_5,s_6},\ee
while for the Corner Transfer matrix
    \be [C_1']_{(s4s5),(s3s6)}=\sum_{s1,s_2,s_7,s_8}
        w_{s_1s_2s_3s_4}[T_1^{s_1}]_{s_5,s_7}[C_1]_{s_7s_8}
        [T_2^{s_2}]_{s_8,s_6}.
        \label{Extension}
    \ee
In these two expressions, the notation $(s_1s_2)$ means that the indices
$s_1$ and $s_2$ are combined to form a single index. In our representation
of the interacting-dimer model, the indices $s_1$ and $s_2$ take initially
5 different values, which implies that $(s_1s_2)$ takes 25 values. At
each iteration of the CTMRG algorithm, the dimension of the vector space
on which act the matrices $C_i$ and $T_i$ is multiplied by a factor of 5.
This exponential growth limits the calculation to small lattice sizes.
To circumvent this limitation, a Singular Value Decomposition (SVD) of
the four extended Corner Transfer matrices $C_i'$ is performed:
    \be C_i'=U_i\Lambda_iV_i^+ \ee
where $U_i$ and $V_i$ are unitary matrices and $\Lambda_i$ is a diagonal
matrix whose elements are the singular values. \correction{The Lapack library
was used to perform this SVD}.
The unitary matrices $U_i$ and $V_i$ are then truncated to a fixed
number of states $\chi$ to keep only the $\chi$ largest singular values.
The change of basis is applied to all matrices. The new Corner Transfer
matrices
    \be C_i''=U_i^+C_i'V_i\ee
are now $\chi\times\chi$ diagonal matrices whose elements are the
largest singular values $\Lambda_i$. This procedure, assimilated to a
renormalization step, minimizes the quadratic error, defined as the
Hilbert-Schmidt norm, between the original Corner Transfer matrix and
its truncated version. The change of basis has to be applied to Transfer
matrices too but differently according to what they are used for. For the
Transfer matrices used to compute the partition function Eq.~\ref{Part},
the change of basis is
    \be [T_1'']^{s_1}=V_4^+[T_1']^{s_1}U_1\ee
while for the Transfer matrices applied to the left (resp. right) of the
Corner Transfer matrices during their extension Eq.~\ref{Extension}
    \be [{T_1^L}'']^{s_1}=U_1^+[{T_1^L}']^{s_1}U_1,\hskip 1truecm
    [{T_2^R}'']^{s_2}=V_1^+[{T_2^R}']^{s_2}V_1.\ee

In section~\ref{sec2}, preliminary calculations are performed with $\chi=25$
to approximatively localize the transition line. Further calculations
are then performed with $\chi=125$ states to improve the location of the
transition line. In section~\ref{sec3}, the critical exponents will be
estimated from computations with $\chi=125$. Note that the truncation
of the matrices introduces systematic deviations of the thermodynamic
averages. In principle, finite Transfer matrices and Corner Transfer
matrices cannot describe a critical system but only gapped systems with a
finite correlation length. In the following, we will consider only
the neighborhood of the transition line and not the critical line itself.

\section{Phase diagram}\label{sec2}
A first series of CTMRG calculations has been performed to determine
the phase diagram of the interacting-dimer model. As discussed above, the
number of states kept during the truncation of the Corner Transfer matrices
is limited in this section to $\chi=25$. Thirteen monomer chemical potentials
have been considered ($\mu_1=-10$, $-2$, $-1$, $-0.4$, 0, $0.16$, $0.26$, $0.28$,
$0.30$, $0.32$, $0.34$, $0.36$ and $0.38$) and 116 temperatures. In the last
part of this section, additional calculations with $\chi=125$ states, giving
more accurate estimates of the critical temperatures, are presented.

\subsection{Order parameters}
\begin{figure}
    \psfrag{M}[Bl][Bl][1][1]{$N$}
    \psfrag{T}[Bl][Bl][1][1]{$T$}
    \centering
    \includegraphics[width=0.60\textwidth]{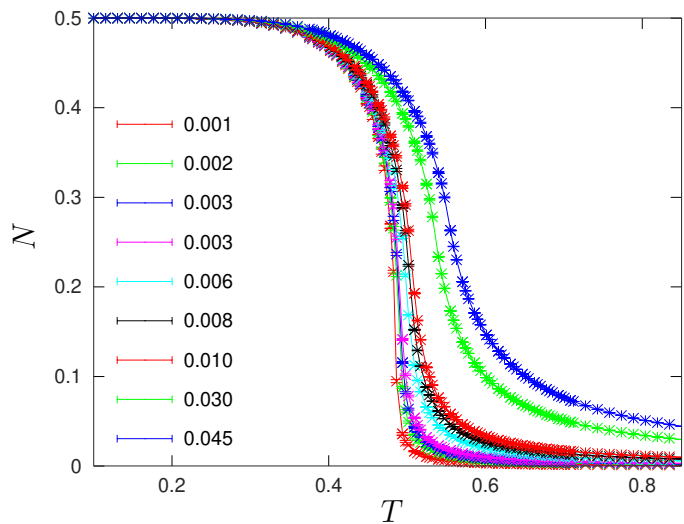}
    \caption{Order parameter $N$ versus temperature for a monomer
    chemical potential $\mu_1=0$. The different curves correspond to
    different chemical potentials $\pm\Delta\mu$ of respectively
    horizontal and vertical dimers ($0.045$ is the curve at the top and
    $0.001$ is at the bottom). The legend gives the values of
    $\Delta\mu$. The data have been computed with $\chi=25$ states.}
    \label{fig7}
\end{figure}

As explained in subsection~\ref{sec1.1}, a field is introduced to break the rotational
symmetry of the model, either by considering a different chemical potential
$\pm\Delta\mu$ for horizontal and vertical dimers or by changing the
interaction to $u\pm\Delta u$ for plaquettes with respectively horizontal
and vertical aligned dimers. The corresponding Boltzmann
weights are given by Eqs.~\ref{PoidsSt2} and \ref{PoidsSt3}.
The linear responses to these symmetry-breaking fields are order
parameters of the transition: $N=\langle (n_h-n_v)\rangle$ where $n_h$
and $n_v$ are the densities of respectively horizontal and vertical dimers
on the central vertex and $P=\langle (p_h-p_v)\rangle$ where
$p_h$ and $p_v$ are the numbers of plaquettes with horizontal and
vertical dimers on the central vertex.
On Fig.~\ref{fig7}, the order parameter $N$ is plotted versus the
temperature $T$ for a monomer chemical potential $\mu_1=0$. The same
curve is obtained for $P$. For both order parameters,
the same results are obtained with a shift $\Delta u$ of plaquette
interaction with horizontal or vertical dimers rather than a dimer
chemical potential $\Delta \mu$. As the monomer chemical potential
$\mu_1$ is increased, the transition becomes steeper and steeper and
the transition temperature decreases.

\subsection{Local entropy}
\begin{figure}
    \psfrag{S}[Bl][Bl][1][1]{$S_{\rm loc}$}
    \psfrag{T}[Bl][Bl][1][1]{$T$}
     \centering
    \includegraphics[width=0.60\textwidth]{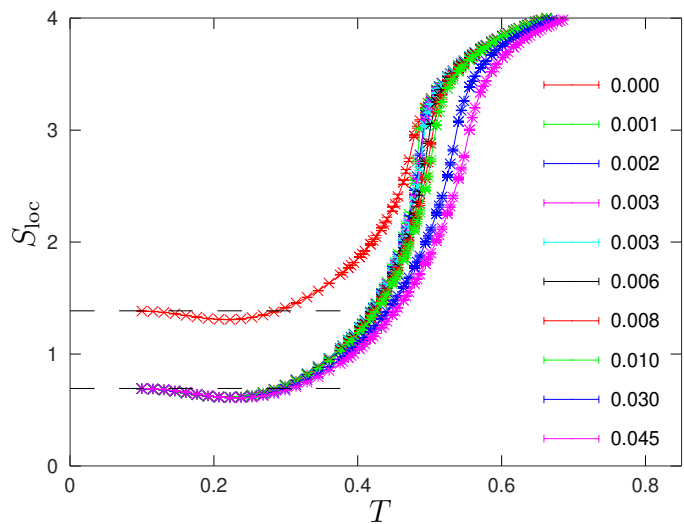}
    \caption{Statistical entropy $S$ on the central vertex
    versus temperature for a monomer chemical potential $\mu_1=0$.
    The different curves correspond to different chemical potentials
    $\pm\Delta\mu$ of respectively horizontal and vertical dimers.
    The values of $\Delta \mu$ are given in the legend. The red
    crosses correspond to $\Delta\mu=0$ while the other points
    correspond to positive symmetry-breaking fields $\Delta\mu$.
    The dashed lines correspond to the constant values $\ln 4$
    and $\ln 2$. The data have been computed with $\chi=25$ states.
    }\label{fig8}
\end{figure}

The entropy $S_{\rm loc}$ on the central vertex is easily computed in
a CTMRG simulation and gives some information on the nature of the
low-temperature phase. The probability distribution on the central
vertex is computed as
    \be \wp(s_1,s_2,s_3,s_4)={1\over{\cal Z}}w_{s_1s_2s_3s_4}
    \Tr\big[T_1^{s_1}C_1T_2^{s_2}C_2T_3^{s_3}C_3T_4^{s_4}C_4\big]\ee
and leads to the statistical entropy
    \be S_{\rm loc}=-\sum_{s_1,s_2,s_3,s_4}\wp\ln\wp.\ee
We emphasize that this local entropy is not the total entropy of the
system. A partial trace over the degrees of freedom described in an
effective way by the Transfer matrices and the Corner Transfer matrices
has been performed. The entropy associated to these degrees of freedom
is therefore lost and $S_{\rm loc}$ is only a lower bound of the total
entropy of the system. $S_{\rm loc}$ can be viewed as the classical
analogue of the quantum entanglement entropy of the central vertex with the
rest of the system. As can be seen on Fig.~\ref{fig8} in the case
of a monomer chemical potential $\mu_1=0$, the local entropy is nicely
compatible with the value $S=\ln 4\simeq 1.387$ in the limit of
zero temperature when no symmetry-breaking field is applied and
with $S=\ln 2\simeq 0.693$ when a different chemical potential
$\pm\Delta\mu$ is assigned to horizontal and vertical dimers.
The same is also observed for a different interaction strength
$\Delta u$ and for all considered values of the monomer chemical
potential $\mu_1$. These values of the local entropy are consistent
with the fact that the ground state of the interacting-dimer model
is four-fold degenerated. The dimer chemical potentials $\pm\Delta\mu$
break rotational symmetry and therefore, reduce the degeneracy of the
ground state to 2. Note that fixing the four indices $s_1$, $s_2$, $s_3$,
and $s_4$ determines completely the ground state. The environment
of the central vertex has no additional degeneracy in the ground state.
As a consequence, the local entropy is equal to the total entropy of
the system at zero temperature.

\subsection{Ratio of the two largest singular values}
\begin{figure}
    \psfrag{gap}[Bl][Bl][1][1]{$g$}
    \psfrag{T}[Bl][Bl][1][1]{$T$}
    \centering
    \includegraphics[width=0.60\textwidth]{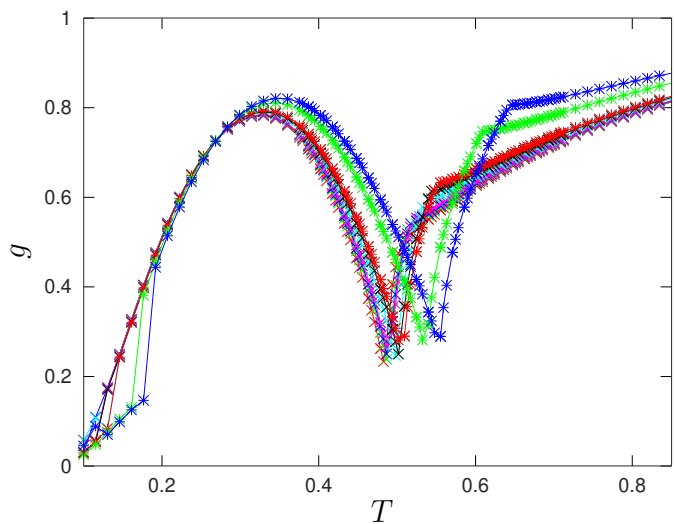}
    \caption{Logarithm of the ratio of the two largest singular values
    of the Corner Transfer matrix (Eq.~\ref{Gap}) versus temperature
    for a monomer chemical potential $\mu_1=0$.
    The different curves correspond to different chemical potentials
    $\pm\Delta\mu$ of respectively horizontal and vertical dimers.
    The same symbols and colors as in Fig.~\ref{fig7} have been used.
    The data have been computed with $\chi=25$ states.}
    \label{fig9}
\end{figure}

To estimate the transition temperatures, we studied the quantity
    \be g=\ln{\Lambda_1\over\Lambda_2}      \label{Gap}\ee
where $\Lambda_1\ge \Lambda_2$ are the two largest singular values
of the Corner Transfer matrix. For the Ising model under a magnetic
field $h$, the two largest singular values tend toward the same value
in the limit $h\rightarrow 0$ for all temperatures $T\le T_c$.
The vanishing of $g$ in the ferromagnetic phase is a consequence of
the existence of a ${\mathbb Z}_2$ symmetry when $h=0$.
The vanishing of $g$ is also observed in the ferromagnetic phase
of the $q$-state clock model for both $q\le 4$ and $q>4$.
In contrast, in the interacting-dimer model, $g$ is non-zero in both
the high and low-temperature phases.
Instead, a dip is observed (Fig.~\ref{fig9}). This is surprising since
we have seen that the entropy takes the expected value $\ln 4$ in the
limit $\Delta\mu\rightarrow 0$ (or $\Delta u\rightarrow 0$), which means
that the tensor product encodes correctly the four-fold degeneracy of the
ground-state. We note that the entanglement entropy of the $7$-mers model
has also been observed to display an unexpected behavior at the two
transitions~\cite{Chatelain}. In this case, as in the case of our
interacting-dimer model, the Corner Transfer matrices are not symmetric.
This can easily be seen for a single vertex. Using Eq.~\ref{InitCTM},
one can find several possible monomer-dimer configurations contributing to
$[C_1]_{32}$ while $[C_1]_{23}$ is equal to zero since the two indices
are not compatible in this order. In contrast, Corner Transfer matrices are
symmetric for the Ising model and their eigenvalues are real and equal to
their singular values. This difference may explain the different behavior
of $g$.

\begin{figure}
    \psfrag{gap}[Bl][Bl][1][1]{$g$}
    \psfrag{T}[Bl][Bl][1][1]{$T$}
    \centering
    \includegraphics[width=0.60\textwidth]{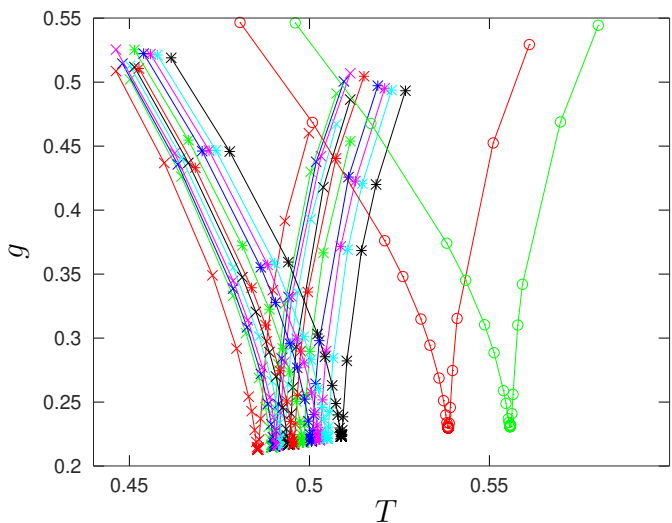}
    \caption{Logarithm of the ratio of the two largest singular values
    of the Corner Transfer matrix (Eq.~\ref{Gap}) versus temperature
    for a monomer chemical potential $\mu_1=0$ with $\chi=125$ states.
    The different curves correspond to different chemical potentials
    $\pm\Delta\mu$ of respectively horizontal and vertical dimers.
    $\Delta\mu=0.001$ for the dip on the left and 0.045 for the one
    on the right.}
    \label{fig10}
\end{figure}

\begin{figure}
    \psfrag{1/Tc}[Bl][Bl][1][1]{$1/T_c(\Delta\mu)$}
    \psfrag{dmu}[Bl][Bl][1][1]{$\Delta\mu$}
    \centering
    \includegraphics[width=0.60\textwidth]{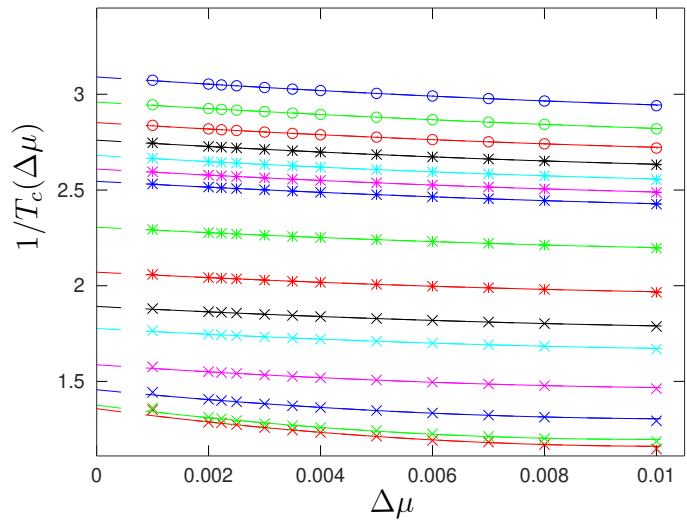}
    \caption{Inverse pseudo-transition temperatures $1/T_c(\Delta\mu)$
    versus $\Delta\mu$. The different curves correspond to different
    monomer chemical potentials $\mu_1=-10$, $-4$, $-2$, $-1$, $-0.4$,
    $-0.2$, $0$, $0.16$, $0.26$, $0.28$, $0.30$, $0.32$, $0.34$, $0.36$
    and $0.38$ (from bottom to top). The continuous curves are quadratic
    fits of the data.}
    \label{fig11}
\end{figure}

\begin{figure}
    \psfrag{T}[Bl][Bl][1][1]{$T_c$}
    \psfrag{mu}[Bl][Bl][1][1]{$\mu_1$}
    \psfrag{du}[Bl][Bl][1][1]{$\Delta u$}
    \psfrag{dmu}[Bl][Bl][1][1]{$\Delta\mu$}
    \psfrag{Jacobsen}[Bl][Bl][1][1]{Alet {\sl et al.}}
    \centering
    \includegraphics[width=0.60\textwidth]{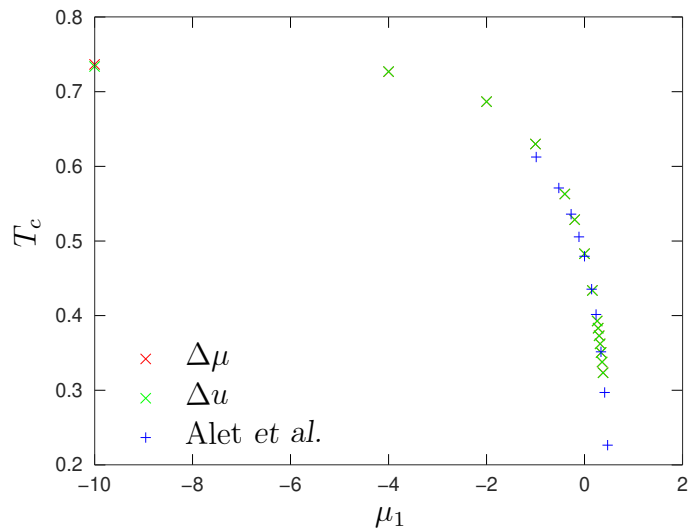}
    \caption{Extrapolated transition temperatures versus the monomer
    chemical potential $\mu_1$. The red crosses have been obtained with
    different chemical potentials $\pm\Delta\mu$ for horizontal and
    vertical dimers while the green ones with different interaction
    strengths $\Delta u$. The blue crosses come from Ref.~\cite{Alet2}.}
    \label{fig12}
\end{figure}

The location of the dip depends on the value of the symmetry-breaking
field, either $\Delta\mu$ or $\Delta u$. It is also more rounded for
negative monomer chemical potentials $\mu_1$ and steeper for large positive
ones. To improve the accuracy, we made additional calculations with
$\chi=125$ states (see Fig.~\ref{fig10}). The temperatures of the dips,
have been determined by dichotomy up to an accuracy of $10^{-5}$
for each monomer chemical potential $\mu_1$ and each symmetry-breaking
field $\Delta\mu$ and $\Delta u$. In the following, these temperatures
are termed as pseudo-critical temperatures $T_c(\Delta\mu)$ and $T_c(\Delta u)$.
We make the conjecture that the critical temperatures of the
model are given by the limit of $T_c(\Delta\mu)$ (resp. $T_c(\Delta u)$)
when $\Delta\mu$ (resp. $\Delta u$) goes to zero. We also expect the following
scaling behavior
    \be|T_c-T_c(\Delta\mu)|\sim \Delta\mu^{y_t/y_h}     \label{EvolT}\ee
where $y_t$ is the temperature scaling dimension and $y_h$ the scaling
dimension of the order parameter $N$. The same scaling relation
is expected for $|T_c-T_c(u)|$. As will be shown in Section~\ref{sec3}, the
two order parameters $N$ and $P$ share the same scaling dimension.
Numerically, a non-linear fit of the data according to the law Eq.~\ref{EvolT}
turned out to be too unstable. We therefore limited ourselves to estimate
the critical temperature $T_c$ by a quadratic fit of the inverse temperature
with the symmetry-breaking field. As can be noticed on Fig.~\ref{fig11},
the fit is good for large chemical potentials $\mu_1$ but not for
$\mu_1=-10$ or $-2$.
% We tried to perform a non-linear fit of the form
%     \be T_c(\Delta\mu)=T_c(0)+\Delta\mu^x\ee
% where the exponent $x$ is expected to be equal to the ratio $y_t/y_h$
% of the thermal exponent $y_t$ and of the order-parameter exponent $y_h$.
% However, the number of points is too small to lead to a stable
% transition temperature at negative monomer chemical potentials.
Our final estimates of the critical temperatures are presented on
Fig.~\ref{fig12}. For large chemical potentials, they are in good
agreement with the transition temperatures given in Table~II of Ref.~\cite{Alet2}.
This agreement provides an {\sl a-posteriori} justification
for the conjecture that the dip of $g$ is located at the transition
point. For negative monomer chemical potentials, the agreement is
not so good because of the difficulty to fit properly $T_c(\Delta\mu)$
or $T_c(\Delta u)$ as \correction{mentioned} above.

\section{Critical behavior}\label{sec3}
In this section, the critical behavior of the interacting-dimer
model along the transition line is studied. The data have been obtained
with $\chi=125$ states.

\subsection{Order parameter exponent}
\begin{figure}
    \psfrag{M}[Bl][Bl][1][1]{$N$}
    \psfrag{dmu}[Bl][Bl][1][1]{$\Delta\mu$}
    \centering
    \includegraphics[width=0.60\textwidth]{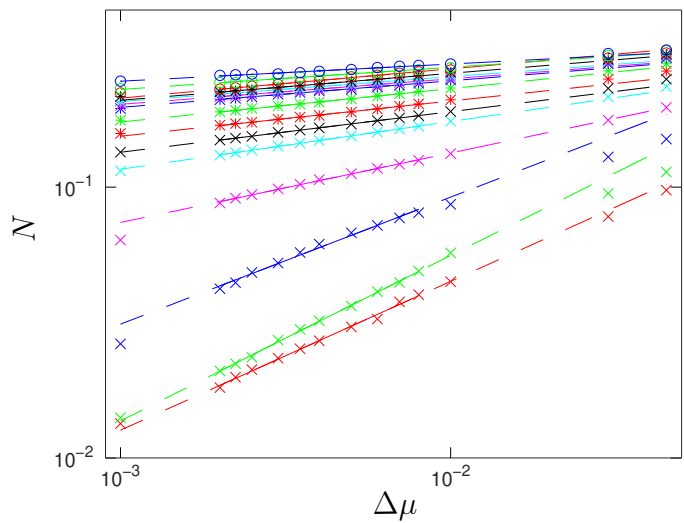}
    \caption{Order parameter $N$ at the pseudo-transition temperatures
    $T_c(\Delta\mu)$ versus $\Delta\mu$. The different curves correspond
    to different monomer chemical potentials $\mu_1$.
    The same symbols and colors as in Fig.~\ref{fig10} have been used.
    The lines are power-law fits of the data. They are represented as
    continuous lines in the window where the data points have been
    fitted and as dashed lines outside.
    }\label{fig13}
\end{figure}

\begin{figure}
    \psfrag{1/delta}[Bl][Bl][1][1]{$1/\delta$}
    \psfrag{Tc}[Bl][Bl][1][1]{$T_c$}
    \psfrag{M,du}[Bl][Bl][1][1]{$N,\Delta u$}
    \psfrag{I,du}[Bl][Bl][1][1]{$P,\Delta u$}
    \psfrag{M,dmu}[Bl][Bl][1][1]{$N,\Delta\mu$}
    \psfrag{I,dmu}[Bl][Bl][1][1]{$P,\Delta\mu$}
    \centering
    \includegraphics[width=0.60\textwidth]{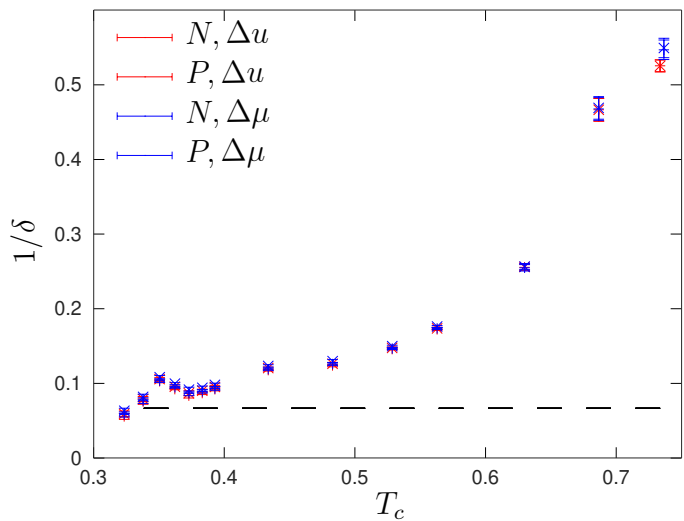}
    \caption{Exponent $1/\delta$ versus the transition temperature for the
    interacting-dimer model. The four sets of data points correspond to the
    order-parameter ($N$ for the symbols $\times$ and $P$ for $+$) from which
    the exponent was extracted and to the field ($\Delta\mu$ in red or
    $\Delta u$ in blue) that were introduced to break the symmetry.
    The dashed line is the expected value $\delta=15$
    for the 4-state Potts model.}
    \label{fig14}
\end{figure}

The two order parameters $N$ and $P$, defined in Sec.~I,
were computed at the pseudo-transition temperatures $T_c(\Delta\mu)$
and $T_c(\Delta u)$ determined as the location of the dip of $g$.
These order parameters are expected to scale in the same way with
$\Delta\mu$ and $\Delta u$:
    \be N\sim \Delta\mu^{1/\delta},\quad
    P\sim \Delta\mu^{1/\delta}      \label{ScalBehav}\ee
where $1/\delta=(d-y_h)/y_h$. The data are presented on Fig.~\ref{fig13}
in the case of the order-parameter $N$. A power-law behavior is
observed in an intermediate range of $\Delta\mu$ and $\Delta u$.
Large $\Delta\mu$ and $\Delta u$ seems to be outside the critical
region where the scaling Eq.~\ref{ScalBehav} holds. For small
$\Delta\mu$ and $\Delta u$, one may suspect that the number of states $\chi$
kept in the truncation of the Corner Transfer matrix becomes too small.
Fig.~\ref{fig13} shows that the exponent $1/\delta$ clearly depends on
the monomer chemical potential.
On Fig.~\ref{fig14}, the exponent $1/\delta$ is plotted versus the
transition temperature $T_c$. The error bars represent the standard
deviation of the fit. They do not take into account the error on the
pseudo-transition temperatures and the systematic deviations due to
the truncation of the Corner Transfer matrices. One can observe that
the estimates obtained by breaking the rotational symmetry with a dimer
chemical potential $\Delta\mu$ are close to those obtained with an
anisotropic interaction $\Delta u$ (compatible within error bars for the
same order parameter) but they are systematically larger.
As in Ref~\cite{Alet2}, we can estimate the temperature of the
tricritical point as the temperature for which the exponent $\delta$
takes the value $\delta=15$ of the 4-state Potts model.
As can been seen on Fig.~\ref{fig13},
our data are compatible with a monomer chemical potential $\mu_1^*\simeq 0.36$
and a transition temperature $T^*=0.34(1)$ at the tricritical point,
slightly above the estimate $0.29(2)$ of Ref~\cite{Alet2}.

\subsection{Thermal critical exponent}
\begin{figure}
    \psfrag{dM}[Bl][Bl][1][1]{${d\over dT}\ln N$}
    \psfrag{dmu}[Bl][Bl][1][1]{$\Delta\mu$}
    \centering
    \includegraphics[width=0.60\textwidth]{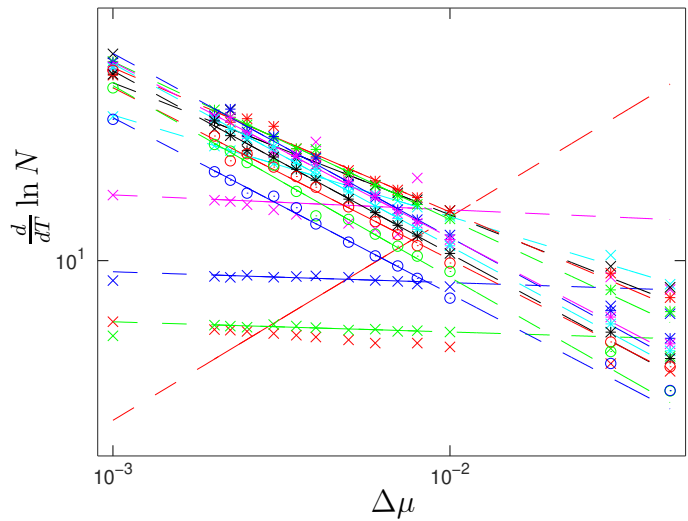}
    \caption{Derivative ${d\over dT}\ln N$ of the logarithm of the
    order parameter at the pseudo-transition temperatures
    $T_c(\Delta\mu)$ versus $\Delta\mu$. The different curves correspond
    to different monomer chemical potentials $\mu_1$.
    The same symbols and colors as in Fig.~\ref{fig10} have been used.
    The lines are power-law fits of the data. They are represented as
    continuous lines in the window where the data points have been
    fitted and as dashed lines outside.
    }\label{fig15}
\end{figure}

In the process of finding the pseudo-critical temperatures $T_c(\Delta\mu)$
and $T_c(\Delta u)$, calculations were performed for at least 24 temperatures
around the dip of $g$. These data were used to compute the temperature
derivative of the order parameter. A linear fit of the two order parameters
$N$ and $P$ with the inverse temperature $\beta=1/k_BT$ was performed in a small
window of width $\Delta\beta=3.10^{-4}$ around the pseudo-critical
temperatures $T_c(\Delta\mu)$ and $T_c(\Delta u)$. The slope of these fits
gives an estimate of the derivatives ${d\over d\beta}N$ and ${d\over d\beta}P$,
from which we then constructed the derivative of the logarithm
    \be {d\over d\beta}\ln N={1\over N}{dN\over d\beta}\ee
that is expected to scale with the symmetry-breaking fields $\Delta\mu$
and $\Delta u$ as
    \be {d\over d\beta}\ln N\sim \Delta\mu^{-\nu_h/\nu},\quad
    {d\over d\beta}\ln P\sim \Delta\mu^{-\nu_h/\nu}      \label{ScalBehavT}\ee
where $\nu_h/\nu=y_t/y_h$. The data are presented on Fig.~\ref{fig15}.
Several points are noticeably outside the curve: a red cross, corresponding
to a monomer chemical potential $\mu_1=-10$, at $\Delta\mu=7.10^{-3}$,
and two pink ones, corresponding to $\mu_1=-1$, at $\Delta\mu=4.10^{-3}$
and $8.10^{-3}$.
These points are due to the fact that sometimes the CTMRG does not find
the true ground state but is trapped in excited states. A step of order
${\cal O}(10^{-3})$ is observed in the curve of the order parameter versus
temperature when the ground state is eventually found. Because of
these steps, our procedure of fitting the curve to estimate the
derivative gives inaccurate estimates of ${d\over d\beta}N$. The latter
then lead to a critical exponent that is completely different from what
is suggested by the rest of the curve and may even be negative as can be
observed on Fig.~\ref{fig15} in the case of the red crosses and the
red line. We did not try to remove these points manually so a few
critical exponents are wrong and do not follow the general tendency.

\begin{figure}
    \psfrag{yt/yh}[Bl][Bl][1][1]{$y_t/y_h$}
    \psfrag{Tc}[Bl][Bl][1][1]{$T_c$}
    \psfrag{M,du}[Bl][Bl][1][1]{$N,\Delta u$}
    \psfrag{I,du}[Bl][Bl][1][1]{$P,\Delta u$}
    \psfrag{M,dmu}[Bl][Bl][1][1]{$N,\Delta\mu$}
    \psfrag{I,dmu}[Bl][Bl][1][1]{$P,\Delta\mu$}
    \centering
    \includegraphics[width=0.60\textwidth]{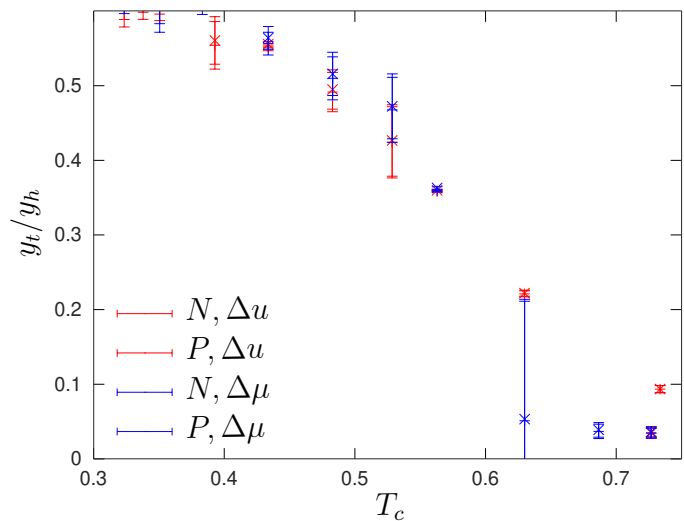}
    \caption{Exponent $1/\nu_h$ versus the transition temperature for the
    interacting-dimer model. The four sets of data points correspond to the
    order-parameter ($N$ for the symbols $\times$ and $P$ for $+$) from which
    the exponent was extracted and to the fields ($\Delta\mu$ in red or
    $\Delta u$ in blue) that were introduced to break the symmetry.}
    \label{fig16}
\end{figure}

\begin{figure}
    \psfrag{yt/yh}[Bl][Bl][1][1]{$y_t/y_h$}
    \psfrag{yh}[Bl][Bl][1][1]{$y_h$}
    \psfrag{M,du}[Bl][Bl][1][1]{$N,\Delta u$}
    \psfrag{I,du}[Bl][Bl][1][1]{$P,\Delta u$}
    \psfrag{M,dmu}[Bl][Bl][1][1]{$N,\Delta\mu$}
    \psfrag{I,dmu}[Bl][Bl][1][1]{$P,\Delta\mu$}
    \centering
    \includegraphics[width=0.60\textwidth]{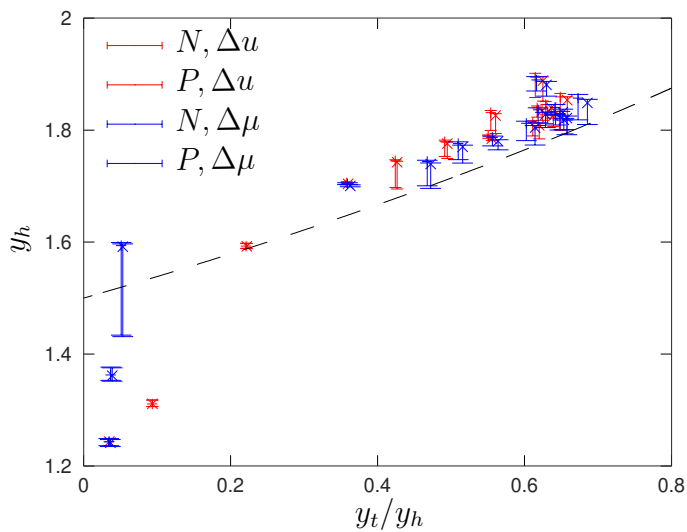}
    \caption{Scaling dimension $y_h$ versus $y_t/y_h$. The four set of data points
    corresponds to the order-parameter ($N$ for the symbols $\times$ and
    $P$ for $+$) from which the exponent was extracted and to the fields
    ($\Delta\mu$ in red or $\Delta u$ in blue) that were introduced
    to break the symmetry.}
    \label{fig17}
\end{figure}

Our final estimates of the critical exponents $y_t/y_h$ are presented
on Fig.~\ref{fig16} versus the transition temperature $T_c$.
The figures~\ref{fig14} and \ref{fig16} do not permit to test the
conjecture that the critical behavior is the same as the Ashkin-Teller
model because the relation between the critical temperatures $T_c(\mu_1)$
and the parameter $y$ is not known. To remove the dependency on $T_c$,
we therefore plotted $y_h$, extracted from $1/\delta$, versus $y_t/y_h$
on Fig~\ref{fig17}. For comparison, the values given by the Ashkin-Teller
exponents Eq.~\ref{ScalDimAT} are plotted on the same figure as a dashed
line from $y=0$ (tricritical point) to $y=3/2$
(limit $\mu_1\rightarrow -\infty$).
Apart from a few points coming from incorrect estimates of the
derivatives and easily recognized by the fact that they correspond
to only one of the four estimates (obtained either from $N$ or $P$ and
with a symmetry-breaking field $\Delta\mu$ or $\Delta u$), our estimates
of the scaling dimensions follow the expected trend, even though they
are systematically slightly above the curve.

\section{Conclusions}
The scaling dimensions of the interacting-dimer model are in agreement with
the conjecture that the critical behavior is the same as the Ashkin-Teller model.
This confirms the ability of the CTMRG algorithm to tackle models with
frustration close to their critical point. The technique has of course numerous
limitations. In contrast to Monte Carlo simulations, only local observables
were considered in this study of the interacting-dimer model. Even though the
computation of non-local observables, as correlation functions, are possible
in principle, it requires extra computations. In the case of the interacting-dimer
model, we have seen that local observables are sufficient to extract the scaling
dimensions $y_h$ and $y_t$ that define completely the critical behavior of the
model. As in all tensor-network algorithms, the accuracy of the CTMRG is limited
by the fact that the truncation of Corner Transfer matrices induces systematic
deviations in the data. We have seen that this limitation can be circumvented,
for the interacting-dimer model, by considering only non-critical points of the
phase diagram. This limitation should also be put in perspective with the
limitation to very small lattices sizes in transfer matrix calculations and
with the critical slowing down that plagues Monte Carlo simulations in the
critical region. Finally, one should mention that the choice of the particular
tensor decomposition of the partition function plays an important role in the
convergence of the CTMRG algorithm. We indeed started this work with a different
decomposition but were unable to reach convergence. It seems that it is
important that the unit cell that is repeated in the ground state, be fully
encoded in the central vertex.

\section*{Acknowledgments}
This work was supported by the french ANR-PRME UNIOPEN grant (ANR-22-CE30-0004-01).
The numerical simulations were performed at the meso-center
eXplor of the universit\'e de Lorraine under the project 2018M4XXX0118.

\section{Bibliography}
\bibliographystyle{cmpj}

\end{document}